\documentclass[aps,preprint,showpacs,showkeys]{revtex4}
\usepackage{graphicx}

\def\pom{$\mathrm{I\!P} \;$}
\def\reg{$\mathrm{I\!R} \;$}

\def\dpe{$\mathrm{D I\!P E}\;$}
\begin{document}
\begin{flushright}
FERMILAB-CONF-16-535-PPD
\end{flushright}
\title{Hadron Spectroscopy in \\ Double Pomeron Exchange Experiments}
\pacs{14.80.Ec,27.75.Dw}
\keywords{pomeron, diffraction, Central exclusive production}

\author{Michael G. Albrow}
\affiliation{Fermi National Accelerator Laboratory,
   Batavia, IL 60510, USA}

\begin{abstract}
Central exclusive production in hadron-hadron collisions at high energies, for example $p + p \rightarrow p + X + p$, where the $+$
represents a large rapidity gap, is a valuable process for spectroscopy of mesonic states $X$. At collider energies the gaps can be
large enough to be dominated by pomeron exchange, and then the quantum numbers of the state $X$ are restricted. Isoscalar $J^{PC} =
0^{++}$ and $2^{++}$ mesons are selected, and our understanding of these spectra is incomplete. In particular, soft pomeron exchanges
favor gluon-dominated states such as glueballs, which are expected in QCD but not yet well established. I will review some published
data.   
\end{abstract} 

\maketitle

\section {Introduction}
While Quantum ChromoDynamics, QCD, is usually called, and generally believed to be,  \emph{THE} theory of strong interactions, it
has only been tested with precision (a) at distances much less that the size of hadrons, 1 fm, corresponding to high momentum
transfers $Q^2 \gg$ 1 GeV$^2$ where perturbation theory applies, or (b) approximating continuous spacetime as a discrete lattice,
as in Lattice QCD. There are many other approaches to modeling hadrons and associated phenomena at large distances, such as bag
models, and string models. Regge theory is based on the sound ideas that scattering amplitudes should obey analyticity, unitarity
and crossing symmetry, together with the $t$-channel exchange of continuous, complex angular momentum. 
In 1958  Pomeranchuk made a prediction from quantum field theory that total cross sections for particles and antiparticles, e.g $\sigma_T(pp)$ and
$\sigma_T(p\bar{p})$, should become equal at very high energy. Total cross sections and elastic scattering are related by the optical theorem, and the
implication is that the $t$-channel exchange in high energy elastic scattering is dominated by an isoscalar and not, e.g., by virtual $\pi$ or $\rho$ exchange,
which dominate at low energy. In Regge theory the former was called the Pomeranchukon, later pomeron, \pom , and the latter are reggeons, \reg.. If the intercept (at $t$ =
0) of the pomeron trajectory $\alpha_t$ were exactly 1.0, these cross sections would not rise with collision energy. 
The discovery at the CERN Intersecting Storage Rings, ISR, that $\sigma_T(pp)$ rises
was strong evidence for the pomeron, with an intercept $\alpha_P(t=0) >$ 1.0.

Without a demonstration that
QCD underlies these large distance phenomena, including confinement of quarks and gluons, we cannot claim that QCD is a
\emph{complete} theory of the strong interaction. Indeed potentially new and unexpected phenomena may be revealed, either
experimentally or through a theoretical breakthrough. Especially interesting is the \emph{QCD vacuum}. It is no exaggeration to
say: ``If we understood the vacuum, we would understand all of fundamental physics". While this obviously applies to 
the Higgs sector at the smallest spacetime scales accessible at the LHC, it should also apply to the $\gtrsim$ 1 fm scale. Any isoscalar-scalar
states with $I^G J^{PC} = 0^+0^{++}$ will be present in the vacuum as fluctuations through Heisenberg's Uncertainty Relation
$\Delta E. \Delta t \gtrsim \hbar$ (as will $e^+e^-$-loops and everything else allowed). 
Such virtual isoscalar states can be ``promoted" to
a real state in a high energy collision of two hadrons (e.g. $pp$ or $p\bar{p}$). We can call this ``diffractive excitation of the
vacuum", or double pomeron, \pom , exchange, \dpe. (For unfamiliar readers, we can simply define the pomeron as the carrier of
the 4-momentum exchanged between two protons scattering elastically at very high energy e.g. at the LHC, in addition to photon exchange,
which dominates at very small scattering angles.) An example of double pomeron exchange
is the reaction $p + p \rightarrow p + X + p$ where the protons are \emph{almost} elastically scattered, carrying an outgoing
momentum fraction $x_F \gtrsim$ 0.95, and the central state $X$ is isolated by rapidity gaps, with no hadrons, $\Delta y \gtrsim 3$
($\Delta y \gtrsim$ 4 or 5 would be better!).  The scattered (anti)proton may diffractively dissociate into a low-mass state ($p^*$) 
provided the longitudinal momentum of $p^*$ has $x_F \gtrsim$ 0.95.

\section{Scalar mesons and glueballs}

I quote from the 2010 Particle Data Group \citep{Nakamura:2010zzi} Note on Scalar Mesons: ``The scalar mesons are especially important to
understand because they have the same quantum numbers as the vacuum ($J^{PC} = 0^{++}$). Therefore they can condense into the
vacuum and break a symmetry such as a global chiral $U(N_f) \times U(N_f)$. The details of how this symmetry breaking is implemented
in Nature is one of the most profound problems in particle physics." But the identification of the scalar mesons is ``a
long-standing puzzle". 

All hadrons are composite, either \emph{baryons} with half-integral spin (fermions) or \emph{mesons} with integral spin (bosons).
Baryons have a $qqq$ valence structure, with sea quark pairs and gluons evolving in with increasing $Q^2$. Mesons have a valence $q\bar{q}$, or in
some cases $q\bar{q}q\bar{q}$, structure, again with a sea of $q\bar{q}$ and $g$ at $Q^2 \neq$ 0. Hybrid mesons are described as having a valence gluon
in addition, such as $q\bar{q}g$, allowing quantum numbers that cannot be just $q\bar{q}$. Fritzch and Gell-Mann already in 1972 \citep{Fritzsch:1972jv} speculated that
there may be meson states that ``would appear to act as if they were made of gluons rather than $q\bar{q}$ pairs". While this can be considered
the first reference to \emph{glueballs}, their paper emphasizes that the authors considered these constituents to be \emph{fictitious}. (This was
about three years \emph{after}
 the deep-inelastic scattering experiments at SLAC and the parton and scaling ideas of Feynman and Bjorken!)  
 Probably this was Gell-Mann's wording; in 1975 Fritzsch and Minkowski published an article \citep{Fritzsch:1975tx} referring to Ref.~\citep{Fritzsch:1972jv} in which they say:
 ``Such states are by definition glue states and constitute a new type of matter. The existence of glue states is a direct consequence 
 of the quark-gluon field theory."
The first paper (in the SLAC data base) with ``Glueball" in the title (now there are 1420) was by D.Robson \citep{Robson:1977pm} in 1977, with ``A Basic Guide for the
Glueball Spotter". He claimed that the scalar mesons form an ideally mixed nonet, ``with an additional scalar, the S$^*$ (now $f_0(980)$) with the
expected properties of a scalar glueball". He also said that the $J^{PC} = 0^{-+}$ $\eta'$(958) contains a large gluon fraction. I cannot possibly do
justice to the many model calculations of glueballs and their properties over the nearly 40 years since then, with still no clear understanding.
Note that the term \emph{gluonium} is  sometimes used to refer to a \emph{gg} state, while glueball is more general. 

\subsection{Bag models and strings}

Bag Models \citep{Konoplich:1981ed,DeTar:1983rw} treat hadrons as bound states of quarks and gluons confined in a fm-size bag with a pressure and surface tension.
Outside the bag is the ``true vacuum", while inside is the ``perturbative vacuum"; a different phase in which quarks and gluons are free. 
There are several forms: the MIT bag model, the SLAC bag model, the soliton bag model, etc. 
Jaffe and Johnson \citep{Jaffe:1975fd} discussed hadrons with unconventional quantum numbers in the bag model, and claimed that the known $J^{PC} =
0^{++}$ mesons in the mass range 600 - 1600 MeV may be members of a nonet of $qq\bar{q}\bar{q}$ rather than P-wave $q\bar{q}$ states.
In 1983 Jezabek and Szwed \citep{Jezabek:1982ic} argued that for glueballs the bag surface fields should be ``TE" (transverse electric, in QCD) which suggests that
the bags have a toroidal topology (one cannot have a transverse field everywhere on the surface of a sphere without sources). 
Interestingly (but I do not know whether the connection is more than a coincidence) in the string models of hadrons glueballs are closed loops of
string, i.e. toroids. In string models quarks and antiquarks are the open ends of directed strings, with the form of an ``I" for a meson and a
``Y" with a three-string junction for a baryon or antibaryon. As glueballs are strings without ends, they have the 
form of an ``O". Their decay occurs by the string-loop 
breaking to generate new ends ($q\bar{q}$) as an excited meson ``I", which in turn breaks to a pair of mesons. The Regge trajectory
$\alpha(t)$ for normal $q\bar{q}$ mesons has a slope $\alpha'$ of order 1 GeV$^{-2}$ linking excited mesons with the same quantum numbers apart from spin.
It is not unnatural for a closed loop of string to have a trajectory with a smaller slope, like that of the pomeron, e.g. 
$\alpha(t) = 1.081 + 0.25$ GeV$^{-2}t$~\citep{Donnachie:1992ny}. A spin $J$ = 2 state would lie on this trajectory (if it is linear) at $M_G(2^{++}) \sim$ 2000 MeV; 
that would be the tensor glueball.
The scalar glueball cannot lie on the pomeron trajectory. 
In this model a barred loop like
$\theta$ would be allowed as a topologically different glueball. Having two three-string junctions this would decay to a baryon-antibaryon pair
(assuming it is not too light, in which case it could be stable).
 Barnes, Close and Monaghan \citep{Barnes:1981kq} calculated order-$\alpha_S$ hyperfine splitting
in the spherical cavity approximation to the MIT bag.
They concluded that the $0^{-+}$ and $2^{++}$ glueball states could be identified with the $\iota$(1440) (now $\eta(1405)$) and $\theta$(1640) (now
$f_2(1640)$, omitted from the PDG summary tables) states, implying that
the lightest scalar $0^{++}$ is around 1000 MeV. This ``may mix with the $S^*(980)$" (now $f_0(980)$).

\subsection{Lattice QCD}

In 1997 Morningstar and Peardon \citep{Morningstar:1999rf} calculated the masses of pure glueballs (in pure SU(3) gauge theory) 
in lattice QCD, in which space and time are treated as discrete. If the lattice spacing is much smaller than the size of a
hadron ($\sim$ 1 fm) this approximation allows (computer-intensive) calculations of hadron masses in terms of one
parameter (a scale with dimension ``mass"). Usually this is done with different lattice spacings $a_s$ and extrapolated to $a_s$ = 0.
Recent developments \citep{Morningstar:1998du} predict the lightest glueballs to have $M_G(0^{++}) = 1710$ MeV and
$M_G(2^{++}) = 2390$ MeV, with uncertainties of about 100 MeV and 125 MeV respectively. Mixing with $q\bar{q}$ states can
affect these masses. 

Table I lists all the established $I$ = 0 and $J^{PC}$ = $0^{++},2^{++}$, and $4^{++}$ states in the 2016 PDG Summary tables \citep{Olive:2016xmw}. These can, in principle, be
produced in \dpe , which is a \emph{quantum number filter} when the 4-momentum transfers $t_1$ and $t_2$ are small. (At larger 
$|t|$ other $J^{PC}$  are allowed, but are suppressed.) 

There is a rather narrow $f'_2(1525)$ state, which decays mostly to $K^+K^-$
and is therefore not a good glueball candidate. For all the states with higher mass the information on the decay modes is very sparse.
The $f_0(1710)$ has only one \emph{established} decay, to
$\eta\eta$; we need to establish its other decays in \dpe production. The nearby \dpe - allowed scalar is $f_0(1500)$,
and possibly both these resonances are mixtures of $q\bar{q}$ and $gg$ states \citep{Ochs:2013vxa,Janowski:2014ppa}. The same lattice QCD calculations predict a
whole spectrum of glueball states with $PC = -+, +-, --$, as well as $M_G(3^{++}) \sim 3600$ MeV, that are not easily 
produced singly in \dpe but can be produced in pairs. Some have masses as high as 4000 MeV, where we do not expect any $q\bar{q}$
mesons. It would clearly be useful to have predictions for the decay modes and widths of these states. Perhaps pair production in
\dpe , and radiative $\Upsilon$ decays, are the best windows on this spectroscopy.

 It is a challenge to measure all these
(sometimes overlapping) states with their decay modes, and partial wave analysis to distinguish $J = 0$ and $J=2$.  
The most favored states for the lightest glueball, albeit mixed with $q\bar{q}$ states, are the scalar $f_0(1500)$ \citep{Ochs:2013vxa} and $f_0(1710)$ \citep{Janowski:2014ppa}.

\begin{table}
\caption{Light meson states allowed in \dpe. Branching fractions are in \%. (PDG 2016)}
\begin{tabular}{||c|c|c|c|c|c|c||}
\hline
Name & M(MeV) & $\Gamma$(MeV) & $I^GJ^{PC}$ & $\pi\pi$ & $K\bar{K}$ & Other modes \\
\hline
$f_0(500)/\sigma$  & 400-550   &  400-700  & $0^+0^{++}$  & $\sim$100 & -  & - \\
$f_0(980)$  & 990$\pm$20 &  10-100  & $0^+0^{++}$  & dominant & seen  & $\gamma\gamma$ seen \\
$f_2(1270)$  & 1275.5$\pm$0.8 &  186.7$^{+2.2}_{-2.5}$3  & $0^+2^{++}$  & $84.2^{+2.9}_{-0.9}$ & $4.6^{+0.5}_{-0.4}$  & 
$4 \pi \sim$ 10\% \\
$f_0(1370)$  & 1200-1500 &  200-500  & $0^+0^{++}$  & seen & seen  & $\rho\rho$ dominant \\

$f_0(1500)$  & 1504$\pm$6 &  109$\pm$7  & $0^+0^{++}$ & 34.9$\pm$2.3 & 8.6$\pm$1.0 & $4\pi$ 49.5$\pm$3.3 \\
$f'_2(1525)$  & 1525$\pm$5 &  73$^{+6}_{-5}$  & $0^+2^{++}$ & 0.8$\pm$0.2 & 88.7$\pm$2.2 & $\eta\eta$ 10.4$\pm$2.2 \\
$f_0(1710)$  & 1723$^{+6}_{-5}$ &  139$\pm$8  & $0^+0^{++}$ & seen & seen & $\eta\eta$ seen \\
$f_2(1950)$  & 1944$\pm$12 &  472$\pm$18  & $0^+2^{++}$ & seen & seen & $\eta\eta$ seen \\
$f_2(2010)$  & 2011$^{+60}_{-80}$  &  202$\pm 60$  & $0^+2^{++}$ & - & seen & $\phi\phi$ seen \\
$f_4(2050)$  & 2018$\pm$11  &  237$\pm 18$  & $0^+4^{++}$ & 17\% & $\sim$0.7\% & $\eta\eta$ 0.2\%  \\
$f_2(2300)$  & 2297$\pm$28 &  149$\pm$40  & $0^+2^{++}$  & - & seen & $\phi\phi$ seen \\
$f_2(2340)$  & 2345$^{+50}_{-40}$ &  322$^{+70}_{-60}$  & $0^+2^{++}$  & -           & - & $\phi\phi, \eta\eta$ seen \\

\hline
\hline
\end{tabular}
\end{table}

\section{Double pomeron exchange: history}

Low and Nussinov proposed in 1975 \citep{Low:1975sv,Nussinov:1975mw} that to lowest order the pomeron, \pom , is a pair of gluons in a color singlet.
This is still considered a very good approximation, and means that double pomeron exchange would be a good reaction to produce glueballs.
Experimental searches for \dpe started already in 1969 \citep{Lipes:1969ed} in the Brookhaven 80" bubble chamber with 25 GeV/$c$ pions: $\pi^- p \rightarrow \pi^- +
(\pi^+\pi^-) + p$. The centre-of-mass energy was very low, $\sqrt{s}$ = 6.9 GeV, and the full rapidity span between the
protons is $\Delta y_{pp} = 2 \times$ ln$\sqrt{s}/M_p$ = 4.0, so it was kinematically impossible to have
two large rapidity gaps. They found 250 events with the two pions between the protons (in rapidity), but they had the characteristics 
of reggeon exchange, not \dpe$\!\!$; 
$\rho$- and $\omega$-reggeon exchanges dominated. 
Later bubble chamber searches~\citep{Derrick:1974fj}, with higher energy beams (205 GeV/$c$), also did not succeed in making an observation of \dpe$\!\!$. 
In 50,000 pictures 191 $ p + p \rightarrow p + (\pi^+\pi^-) + p$ events were selected, showed no evidence for \dpe , and gave an
upper limit on the cross sections $\sigma_{DPE} < 44 \; \mu$b. In 1975 a France-Soviet Union collaboration ~\citep{Denegri:1975pb} used a 69 GeV/$c$ beam
on a liquid hydrogen target. The events were all compatible with single diffractive dissociation (one large rapidity gap, not two),
and they quoted an upper limit $\sigma$(\dpe ) $<$ 20 $\mu$b. 

The first observations of \dpe came in 1976 at the CERN ISR, but before discussing those we should mention the last ``heroic"
attempt to do this physics with a hydrogen bubble chamber. In 1980 Brick \emph{et al.}~\citep{Brick:1980jf} took 500,000 photographs with 147 GeV/$c$ $\pi$, K, and p beams,
finding just 47 \dpe candidates corresponding to a cross section $\sigma \sim$ 20 - 50 $\mu$b. The conclusion is that the study of \dpe requires the higher
$\sqrt{s}$ of colliding beams, and electronic detectors, not bubble chambers. However at the CERN Omega-spectrometer, a major fixed target facility,
many studies were done with beams up to 450 GeV/$c$, $\sqrt{s}$ = 29 GeV. The full rapidity span between target and beam is 
$\Delta y = 2 \times$ ln$\sqrt{s}/M_p$ = 6.86, which is on the threshold of allowing two rapidity gaps of 3 units with a central low-mass state.
I return to the Omega experiments, after discussing the CERN Intersecting Storage Rings, ISR.

The ISR started producing $pp$ collisions in 1971 at much higher energies than any fixed target
experiments (even today, being equivalent to 2.1 TeV/$c$ protons on a hydrogen target). 
The first ``evidence paper" for \dpe by Baksay \emph{et al.} \citep{Baksay:1975dc} was from a relatively simple experiment with no magnetic field. 
Small-angle protons above and below the outgoing
beam pipes were tracked in proportional chambers, a cylindrical scintillator hodoscope covered the central region and two counter hits were required, and in-between veto counters
established pseudorapidity ($\eta$) gaps of at least 2 units. To avoid elastic scatters the protons were in the same azimuthal direction, ``UP + UP" or ``DOWN + DOWN". 
Data were taken at several ISR beam energies from 15 + 15 GeV to 31 + 31 GeV; \dpe cross sections were in the range 16 - 28 $\mu$b, less than $10^{-3} \times
\sigma_{inel}$. Although the proton momenta were not measured, assuming they had the beam momentum (a good approximation)
the $t$-slope is $b = -9.9 \pm$ 1.8 GeV$^{-2}$, compatible with half the elastic slope as expected.

The only attempt at a large solid angle detector with tracking in the early days at the ISR was the Split Field Magnet (SFM) facility.
This had a dipole field in the forward directions,
  but in the central region the acceptance and the magnetic field were complicated.
  Leading protons could be well measured, a fact exploited by experiment R407/408, which also observed \dpe in 1976~\citep{DellaNegra:1976coo}. 
  Figure 1(left) shows cross sections vs. $s$, fit \citep{Desai:1978rh} to a falling reggeon, \reg, component and a rising \pom component, which 
dominates only for $\sqrt{s} \gtrsim$ 50 GeV. 
If one fixes the central state to have $|y(\pi\pi)| < 1$, as $\sqrt{s}$ increases the gaps get bigger and the cross section decreases.
If one instead fixes two rapidity gaps between the protons and the central pions $\Delta y \geq$ 3, the central region expands with $\sqrt{s}$ and the cross section rises.

\begin{figure}
\includegraphics[width=\textwidth]{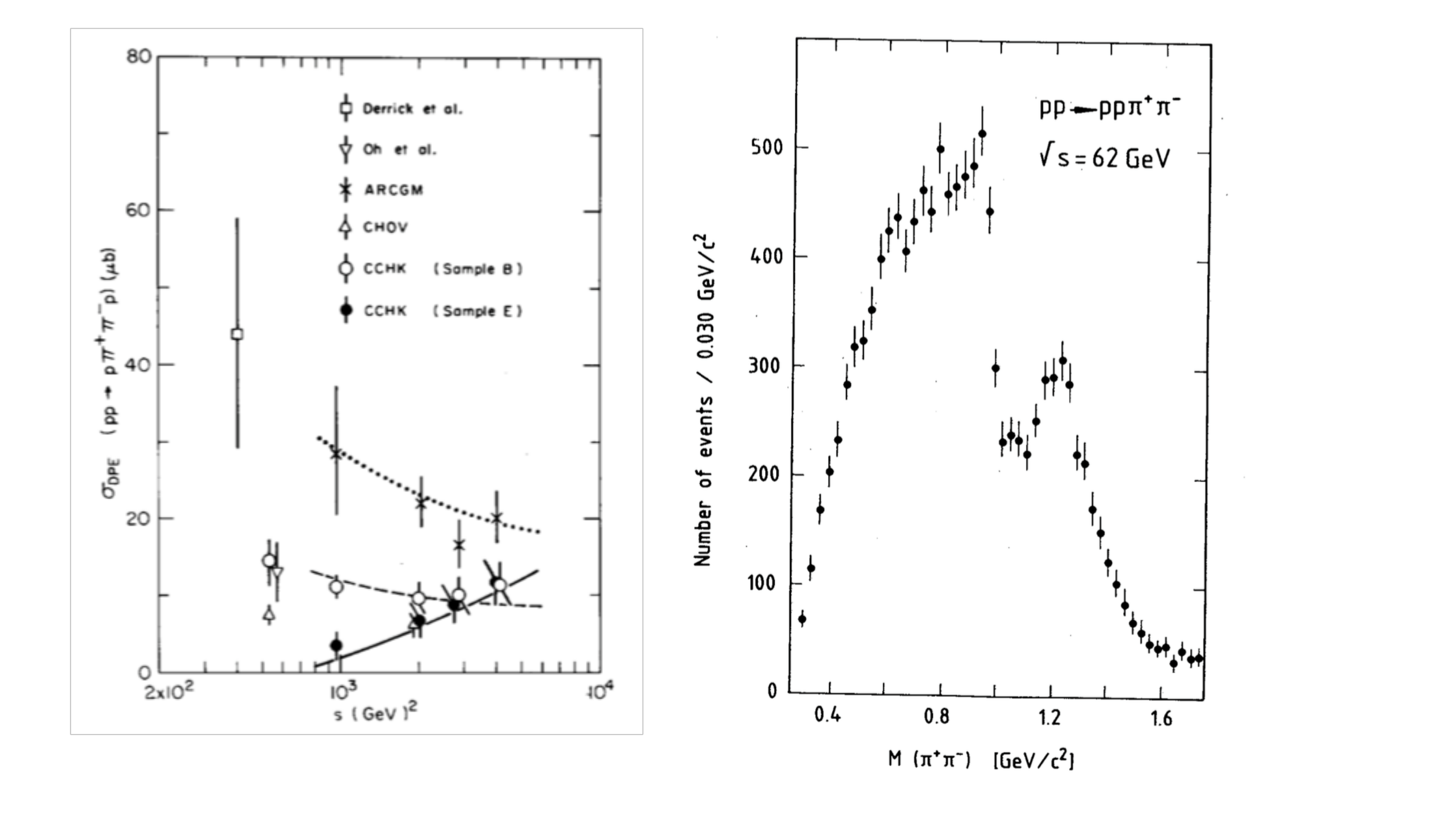}
\caption{Left: Experimental results on $\sigma_{DPE}$ up to ISR energies, with Regge fits \citep{Desai:1978rh}.
The full circles and the rising solid line are for two gaps with $\Delta y > 3$. The dashed line is for $|y_\pi| <$ 1.0 and the dotted line
for $|y_\pi| <$ 1.6.
Right: The central exclusive $\pi^+\pi^-$ invariant mass distribution at $\sqrt{s}$ = 62 GeV (SFM \citep{Breakstone:1986xd}).}
\end{figure}

  After the ISR was shut down in 1984 Breakstone \emph{et al.}~\citep{Breakstone:1986xd,Breakstone:1986mz,Fischer:2014eta} published a more detailed SFM study of a 4-C fit to 
  $p + \; \pi^+\pi^- + p$ with $x_F(p) >$ 0.9 (protons and pions are assumed, but not identified).             
  They found $t_1$ and $t_2$ to be uncorrelated, and to have an exponential slope $b$ = $-$6.1 GeV$^{-2}$, half the elastic slope, for both $t_1, t_2$, and $(t_1+t_2)$,
  as expected for \dpe$\!\!$. The cross section is about 10 $\mu$b, showing some rise through this energy range~\citep{Albrow:2010yb}. The $M(\pi\pi)$ spectrum rises from threshold
  up to 1000 MeV, with no sign of a $\rho$-meson (forbidden in \dpe$\!\!$), and then drops rapidly, see Figure 1(right). This behavior is called a ``cusp", occuring when the $K\bar{K}$
  threshold opens, but the narrow $f_0(980)$ meson occurs at nearly the same mass. A bump in the cross section looks like the $f_0(1270)$ state, but a partial wave
  analysis showed that the $J = 2$ D-wave is dominated there by $J = 0$ S-wave. This raised the suggestion \citep{Minkowski:2002nf} that the data all the way up to 1500 MeV, where there is a
  break, may be dominated by the $f_0(500)$/$\sigma$, a very broad ($\Gamma$ = (400 - 700) MeV) $I^G J^{PC} = 0^+0^{++}$ (poorly understood) state, destructively
  interfering with the $f_0(980)$ to form a dip.

  The Axial Field Spectrometer, AFS, was designed for high-$E_T$ jet physics, with a uranium-scintillator calorimeter covering $\Delta \phi = 2 \pi$. 
  To search for glueballs in \dpe$\!\!$, sets of drift chambers for proton tracking were added~\citep{Akesson:1983jz,Akesson:1985rn} along the beam pipes,
  with veto counters covering $1.5 < |\eta| < 3$. Events kinematically compatible with $p + h^+h^- + p$ with $x_p >$ 0.95 were selected, and the central hadrons were
  identified at low momenta by ionization, $dE/dx$. At $\sqrt{s}$ = 63 GeV there were 87,000 $\pi^+\pi^-$, 523 $K^+K^-$, and 64 $p\bar{p}$ events, with a small amount
  of data also at $\sqrt{s}$ = 45 GeV. The general features are similar to those in Figure 1(right), including S-wave dominance up to about 1500 MeV,
  apart from a small $f_0(1270)$. The only established~\citep{Agashe:2014kda} scalar meson in this region is the broad $f_0(1370)$. The data extend to 3500
  MeV, showing a broad bump from 1500 to 2500 MeV. 
  
 The ISR also provided $\alpha-\alpha$ collisions at $\sqrt{s}$ = 126 GeV, and both the AFS~\citep{Akesson:1985rn} and the CERN-Naples-Pisa-Stony 
 Brook experiment~\citep{Cavasinni:1985bi} measured $\alpha + \pi^+\pi^- + \alpha$ events, clearly coherent, as the $\alpha$ stay intact while pions are created. The mass spectrum has the
  same shape as in $pp$, within the large statistical uncertainty, the $t$-slope is about half that of elastic $\alpha\alpha$ scattering, and $\sigma$(\dpe$\!\!$) is
  about a factor of two higher than in $pp$ collisions.

 \begin{figure}[!ht] 
 \begin{center}
\includegraphics[width=0.55\textwidth,angle=270]{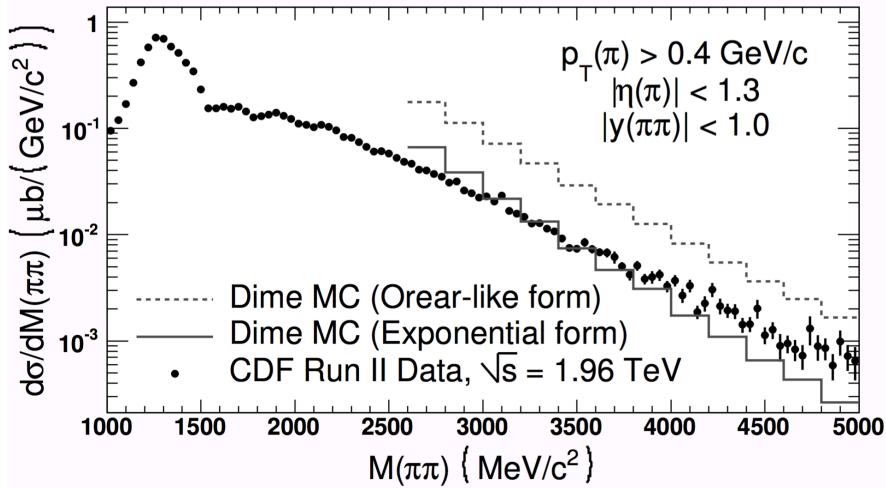}
 \caption{The cross section for exclusive $\pi^+\pi^-$ vs $M(\pi\pi)$ for all $p_T(\pi\pi)$, assuming the hadrons to be pions,
 at $\sqrt{s}$ = 1960 GeV, above the rapid drop at 1000 MeV/$c^2$. The data are compared to the \textsc{dime} Monte
 Carlo~\citep{Harland-Lang:2013dia}. Figure from Ref~\citep{Albrow:2014ita}.}
 \end{center}
\end{figure}

In the post-ISR years many excellent studies of central exclusive hadron production on a fixed target 
were done with $\sqrt{s}$ = 13 - 29 GeV using the Omega-spectrometer at CERN, with many different central states~\citep{Kirk:2014nwa}. 
But some \reg + \pom backgrounds were always present. The last fixed target \dpe experiment
was E690~\citep{Gutierrez:2014yqa} at the Fermilab Tevatron with an 800 GeV/$c$ proton beam ($\sqrt{s} =$ 40 GeV). Exclusive $X = \pi^+\pi^-, K^0_s K^0_s, 
K^0_s K^\pm \pi^\mp$ and $\phi\phi$ channels were studied. The slow recoil proton was inferred from the missing-mass-squared of the event ($M^2_{miss}
\sim m_p^2$). A partial wave analysis (PWA) was made to select $S$-wave ($J$ = 0) and $D$-wave ($J$ = 2) intensities. The S-wave $\pi^+\pi^-$ spectrum
shape up to 2000 MeV is essentially identical to that measured earlier at the ISR, with only a small D-wave $f_2(1270)$. However, if the
fast proton has $p_T >$ 1 GeV/$c$ the $f_2(1270)$ becomes more prominent. 

Very little \dpe data was taken at the CERN $Sp\bar{p}S$, but papers from experiments UA1~\citep{Joyce:1993ru} and UA8~\citep{Brandt:2002qr} are 
discussed in Ref.~\citep{Albrow:2014ita}. 

The Tevatron $p\bar{p}$ collider with $\sqrt{s}$ = 1960 GeV was the perfect machine for \dpe ; at $\sqrt{s}$ = 1960 GeV we have $y_{BEAM}$ = 7.64.
With a good solenoidal central detector, with charged hadron identification by time-of-flight, as in CDF, one can have 
rapidity gaps $>$ 5 units adjacent to the central hadron-pair, providing
essentially pure pomeron exchange. (Photon exchange can also give such large gaps, and both $\gamma$ \pom and $\gamma\gamma$ events were observed when the central
state is forbidden in \dpe, as for $e^+e^-$\citep{Abulencia:2006nb}, $\mu^+\mu^-$\citep{Aaltonen:2009kg, Aaltonen:2009cj}, 
and $J/\psi$\citep{Aaltonen:2009kg}). Unfortunately an early installation of Roman pots to measure elastic scattering and single
diffraction was not retained long enough to study \dpe with both protons detected. However sets of scintillator paddles (Beam Shower Counters, BSC) were installed around the beam pipes,
and could be used as rapidity gap detectors and also for triggering. In the last months before the Tevatron shut down (30 September 2011) CDF recorded 10$^8$ 
events with two forward rapidity gaps, and two or more charged particles in the central region~\citep{Albrow:2014ita}. Of these, 127,340 events had a pair of 
hadrons with $Q$ = 0, $p_T >$ 0.4 GeV/$c$, and $|\eta| < 1.0$ 
and nothing else detected in $|\eta| <$ 5.9. When the $p_T$ of the pair is large enough to have acceptance for low-mass pairs, 
the $f_0(980)$ is seen as a small peak followed by a sharp
drop, as typical in these spectra, and it is followed by a large peak, see Figure 2, which is probably both $f_2(1270)$ and $f_0(1370)$, although unfortunately the spin states
could not be separated. (The $X \rightarrow \pi^+\pi^-$ decay is consistent with being isotropic up to 1500 MeV, within the limited angular coverage.)

In the last two years new results on spectroscopy in \dpe have come from RHIC ($pp$-collisions at $\sqrt{s}$ = 200 GeV) and 
CMS at the LHC ($pp$-collisions at $\sqrt{s}$ = 7 TeV). These are
reported at this workshop by Sikora 
and Khakzad \citep{Khakzad:2016avq}, and I shall not discuss them here. Both experiments have a great
deal more data already recorded, and we can look forward to seeing the spectra with higher statistics. There are also data from CDF on channels other than the
published $\pi^+\pi^-$ that are still being analysed. Some recent theoretical calculations for these processes can be found in Refs.\citep{Harland-Lang:2013dia}
and \citep{Lebiedowicz:2016ioh}.

What is really needed to make a leap forward is
 high statistics (e.g. 10$^6$ events/channel) 
   with both protons measured at high $\sqrt{s}$, at RHIC or the LHC, and in many channels with identified hadrons including (but 
   not only) $K^+K^-$, $K^0_S K^0_S$, $\phi \phi$, $\eta \eta, \eta \eta', \eta' \eta', K^0_S K^\pm \pi^\mp,$ and $ \pi\pi KK$.  
   It may be that the \dpe spectra are different when the protons are detected at
small $|t|$ than when only gaps are required; this could now be tested directly in CMS-TOTEM low-pileup runs at the LHC, by comparing central states 
with leading protons and with leading showers in the Forward Shower Counters, FSC.
   This could actually be done in a few days of low pileup running at the LHC with special triggers. 
   If there is, as expected in QCD,  a scalar glueball with mass $>$ 1000 MeV it will probably be quite wide and therefore 
   have such a short lifetime that if produced \emph{inclusively} it will decay within the hadron formation region, $\sim$ 1 fm.  It will not be an 
   isolated hadron, but live and die in a ``messy" environment. Only in direct \dpe production (or perhaps also 
   in $e^+e^- \rightarrow \Upsilon \rightarrow \gamma + X$) can single glueballs be alone, in a clean 
   (in fact, vacuum) environment. More than 40 years after glueballs were proposed it is high time we understood the 
   isoscalar $J^{PC} = 0^{++}$ and $2^{++}$ spectra, and the QCD vacuum at distance scales $\sim$ 1 fm.

\medskip

\bibliography{diff2016}

\end{document}